\documentclass[showpacs,preprintnumbers,amsmath,amssymb,12pt,pre,aps]{revtex4}
\usepackage{amsmath,amssymb}
\usepackage{epsfig}
\usepackage{citesort}
\usepackage{graphicx}
\usepackage{times}
\begin{document}

\title{Velocity Distribution and the Effect of Wall Roughness 
in Granular Poiseuille Flow}
\author{K. C. Vijayakumar and Meheboob Alam\footnote{
Author to whom correspondence should be addressed:
Electronic Address: meheboob@jncasr.ac.in; Published in \\ {\it Physical Review E}, vol. 75, 051306 (2007)
}
}
\affiliation{
Engineering Mechanics Unit, Jawaharlal Nehru Center for Advanced
Scientific Research, Jakkur PO, Bangalore 560064, India
}

\date{\today}

\begin{abstract}
From  event-driven simulations of a gravity-driven channel flow of inelastic hard-disks,
we show  that the velocity distribution function remains close to
a Gaussian for a wide range densities (even when the Knudsen number 
is of order one) if the walls are smooth and 
the particle collisions are nearly elastic. 
For dense flows, a transition from a Gaussian to a power-law
distribution for the high velocity tails occurs with increasing dissipation 
in the center of the channel, irrespective of wall-roughness. 
For a rough wall, the near-wall distribution functions are
distinctly different from those in the bulk even in the quasielastic limit.
\end{abstract}


\maketitle

Granular materials, a collection of macroscopic particles, are
important in many chemical and pharmaceutical industries 
as well as in geophysical contexts (avalanche, sand dune, etc.).
In the rapid flow regime~\cite{Kada99}, the theory 
for flowing granular materials is largely based on the dense gas kinetic 
theory that incorporates the {\it inelastic}  nature of
particle collisions. At the heart of such gas- or liquid-state continuum 
theories lies the concept of `coarse-graining' over distribution functions
while making a transition from the  particle-level 
properties to the macro-scale fields. 
Unlike the molecular fluid for which 
the Maxwell-Boltzmann (Gaussian/Maxwellian) distribution plays the role of 
the `equilibrium' distribution function, however, the granular fluid does 
not possess any `equilibrium' state~\cite{Kada99,SG98} due to the microscopic 
dissipation of particle collisions. However, there are `non-equilibrium' 
(driven) steady-states for various canonical granular flow configurations 
for which the Gaussian distribution is the leading-order 
velocity distribution~\cite{SG98} in appropriate limits. 
A systematic study of distribution functions is, therefore, of interest from the viewpoint of 
developing constitutive models for granular flows as well as to pinpoint the 
range of validity of any adopted theory.
Another important issue that needs attention is the derivation of continuum boundary
conditions for granular flows~\cite{Richman88} where it is generally assumed
that the distribution function in the near wall-region is the same as that in the bulk
which is unlikely to hold as we shall show here.

In `driven' granular flows, the deviation of velocity distribution from 
a Gaussian has been studied through theory~\cite{SG98,NE98}, simulation
~\cite{TT95,CLH00} and experiment~\cite{RM00,KH00,MN05}. 
In  Ref.~\cite{TT95} it has been shown that the velocity 
fluctuations in a vibrated bed of particles follow a Gaussian distribution 
in the solid phase and a power-law distribution (with 
an exponent $-3$) in the fluidized phase. For the plane shear 
flow~\cite{SG98}, the velocity distribution function is well fitted by 
an exponent of a second-order polynomial in the norm of the fluctuating 
velocities with angle-dependent coefficients.
Theoretical works~\cite{NE98} for a randomly  heated granular gas, 
based on the Boltzmann-Enskog  equation, have predicted velocity distribution 
functions of the form $P(v)\sim\exp\left(-\gamma v^{\alpha}\right)$, with the 
exponent of high-velocity tails being $\alpha=3/2$ which also depends on the 
level of inelastic dissipation. Experiments~\cite{RM00} for a granular gas 
confined between two vertical plates and driven into a steady state via vertical 
vibrations have shown that $\alpha\sim 1.55\pm 0.1$, for a wide range of 
frequency and amplitude of vibrations. Some recent experiments~\cite{KH00},  
however, showed that the high-velocity tails cannot be 
described by a `single' universal exponent.

In this paper we consider the `granular' Poiseuille flow which is the 
gravity-driven flow of granular materials through a two-dimensional 
channel~\cite{FC06}, focusing on the `rapid' flow regime~\cite{Kada99}.
The simulated system is a channel of length $L$ along the periodic 
$x$-direction and bounded by two plane solid walls, parallel to the 
$x$-direction, with a separation of width $W$ (along the $y$-direction). 
The granular material, consisting of $N$ identical 
rigid and smooth disks of equal mass and diameter $d$, is driven by 
gravity along the $x$-direction. The interactions that are allowed 
are instantaneous `dissipative' collisions between pairs of particles 
and between a particle and the  walls,
via an event-driven  algorithm~\cite{Lubachev91}. 
The dissipative nature of particle collisions is characterized by the 
coefficient of normal restitution, $e_n$, which is the ratio between the 
pre- and post-collisional relative velocities of the colliding particles.
There is no relative tangential velocity since the particles are assumed 
to be smooth. The solid walls are modeled as frictional surfaces, and a 
particle colliding with a wall is analogous to a particle colliding with 
a particle of infinite mass moving at the velocity of the wall. The 
frictional properties of the walls are modeled using a single 
parameter,  the {\it coefficient of tangential restitution} 
for particle-wall collisions ($\beta_w$), 
which is defined as the fraction of relative tangential momentum 
transmitted from a particle to the wall during a particle-wall collision.
The wall-roughness is  controlled by choosing a specific value of $\beta_w$:
while $\beta_w=1$ corresponds to a fully {\it smooth} wall, $\beta_w=0$
corresponds to a fully {\it rough} wall for which 
the dissipation of energy at walls is maximum and there is no relative 
tangential slip between the particle surface and the wall upon a 
wall-particle collision.

Apart from the wall-roughness parameter $\beta_w$, 
the granular Poiseuille flow is governed by three 
control parameters: the average  volume fraction ($\nu$), the coefficient 
of normal restitution ($e_n$) and the channel width ($W/d$).
It may be noted that the gravitational acceleration ($g$) does not appear
explicitly as a control parameter since it is used as a reference scale
for time ($\sqrt{W/g}$), velocity ($\sqrt{Wg}$) and other mean fields.
In the present simulation, we have
fixed  $N=900$ and $W/d=31$ and varied the channel length 
($L/W$) to change the average volume fraction, 
\[
   \nu
    ={\pi N}/4\left({L/W}\right)\left({W/d}\right)^2.
\]
(The robustness of reported results was checked by increasing the
number of particles by four-fold.)
The system is initially allowed to attain a statistically steady state for which
the stream-wise velocity ($U_x$), volume fraction ($\nu$) and granular 
temperature ($T$), remain invariant in time, but have spatial variations 
along the wall-normal direction ($y$).
All the statistics presented in this paper are computed `bin-wise'
by dividing the channel into $18-$bins along the wall normal direction,
and  collecting the data in each bin over
about $300000$ collisions per particle after reaching the steady state. 
An increase in the number of bins to $31$ did not alter the results; 
a few bins are indicated by arrows in Fig.~1(b),
with $bin=1,18$ being located adjacent to the walls  and 
$bin=9,10$ at the center of the channel.
It is to be noted that $u_x=c_x - U_x(t)$ and $u_y=c_y$ are the particle fluctuating velocity
in $x-$ and $y-$directions, respectively, over the instantaneous mean velocity;
here $U_x(t)$ and $U_x=<U_x(t)>$ are the `instantaneous' mean and `time-averaged' mean
streamwise velocity, respectively.

\begin{figure}[h!]
\begin{minipage}[h!]{8.5cm}
\includegraphics[width=0.95\textwidth]{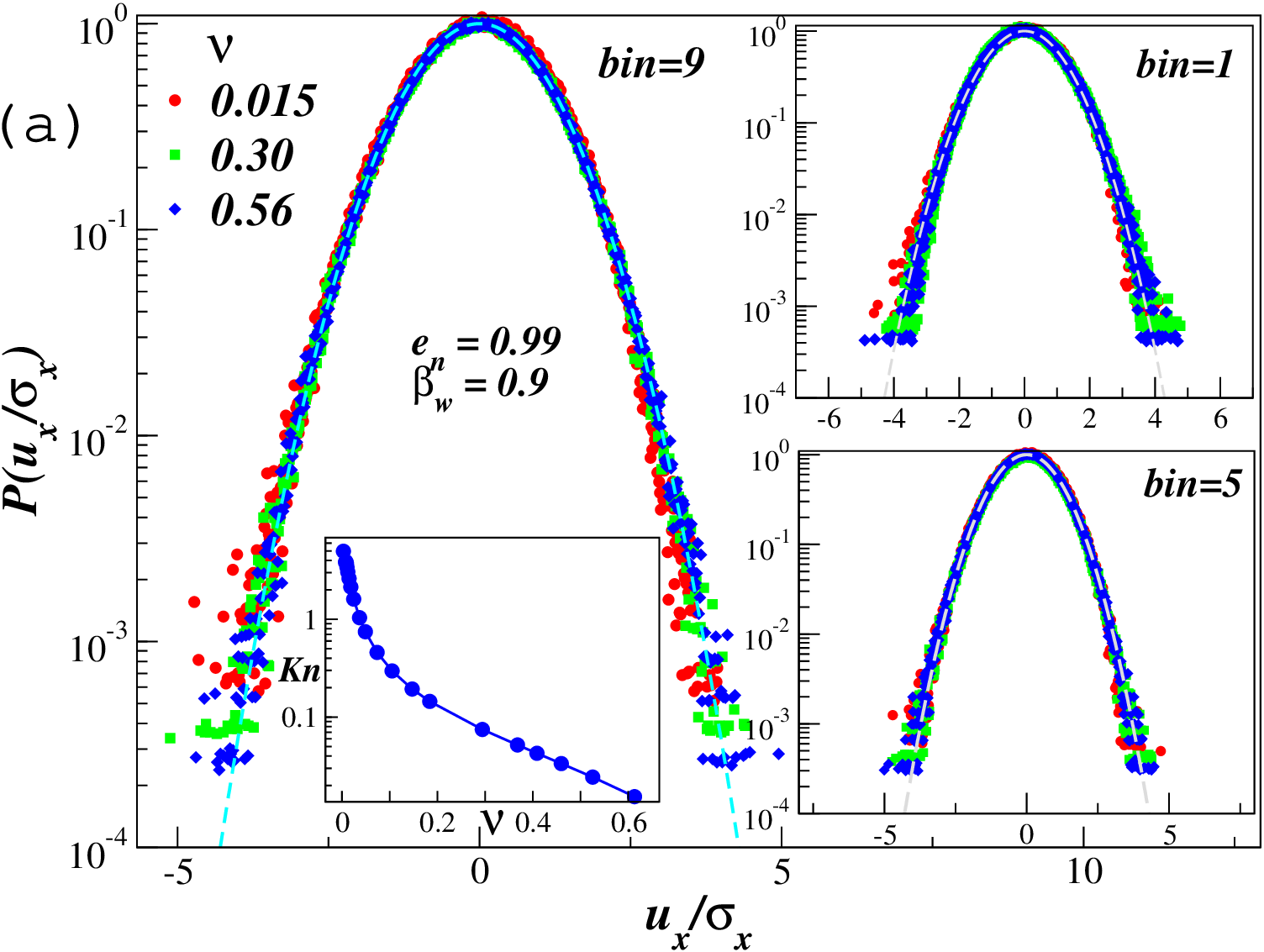}
\end{minipage}
\quad 
\begin{minipage}[h!]{8.5cm}
\includegraphics[width=0.95\textwidth]{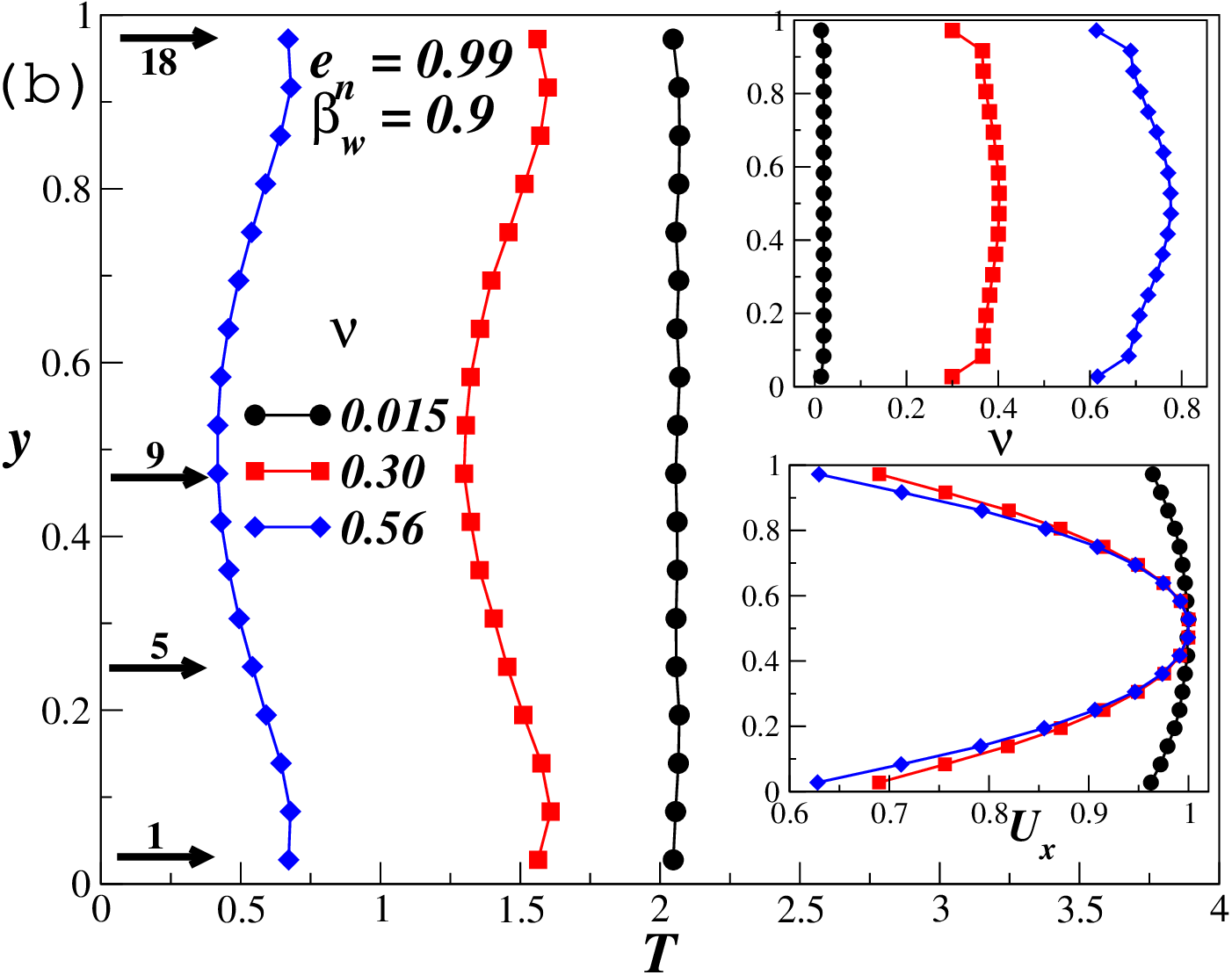}
\end{minipage}
\caption{\small 
(Color online)
(a) Distribution function of  $u_x$ for a range of volume fractions
in the quasi-elastic limit ($e_n=0.99$) for smooth ($\beta_w=0.9$) walls in
different bins. The dashed  curve indicates a Gaussian. 
Left inset shows the variation of Knudsen number, $Kn$, with volume fraction.
(b) The mean velocity ($U_x$), granular temperature ($T$) and volume fraction
($\nu$) profiles across the width of the channel for $e_n=0.99$
and $\beta_w=0.9$ at different volume fractions.  
The arrows near the left-ordinate indicate the locations of some bins.
}
\label{fig1}
\end{figure}

Figure~\ref{fig1}(a) shows the probability distribution function 
of the fluctuating streamwise velocity ($u_x$) for dilute-to-dense 
flows ($0.015\leq \nu\leq 0.56$) in different bins. 
The wall-roughness has been set to $\beta_w=0.9$ for smooth walls, and the 
restitution coefficient to $e_n=0.99$ for {\it quasi-elastic} particle collisions. 
(The distribution function of the fluctuating transverse velocity, $u_y$, looks 
similar, and hence not shown.) 
The local (bin-wise) distribution functions on only one-side of the channel-centerline  are
presented as the distributions on the other side is the same; however, in some 
cases, the distributions on both sides are presented when they differ.
Note that the horizontal axis in the velocity distribution 
plots is scaled by $\sigma_i=\sqrt{\left<u_i^2\right>}$, where the index $i$ denotes
the coordinate direction, and the vertical axis has been scaled such that $P(0)=1$. 
It is remarkable that the velocity distribution function in all bins 
remains a Gaussian for a wide range of densities ($\nu< 0.6$).
This is a surprising result, especially in the dilute limit, since the Knudsen 
number [see left inset in Fig.~\ref{fig1}(a)], which is the ratio between 
the mean free path and the channel width, $Kn=l_f/W$, increases with decreasing 
$\nu$ and becomes of $O(1)$ in the dilute limit, signalling the onset of rarefied flow. 
Even in this rarefied regime, the velocity distribution function remains 
a Gaussian in granular Poiseuille flow with smooth walls. 
From the  profiles of temperature ($T$), mean velocity ($U_x$) 
and volume fraction ($\nu$) in Fig.~\ref{fig1}(b),
we observe that the mean field quantities develop considerable gradients along $y-$direction
with increasing density (and this is more pronounced for $U_x$, 
see lower inset in Fig. 1b), however, they are almost constant over the width of a bin. 
The mean-field gradients do not seem to play  
any role in determining the form of `local' velocity distribution functions
as long as the walls are smooth and the particle collisions are quasielastic.

\begin{figure}[h]
\begin{minipage}[h]{8.5cm}
\includegraphics[width=0.95\textwidth]{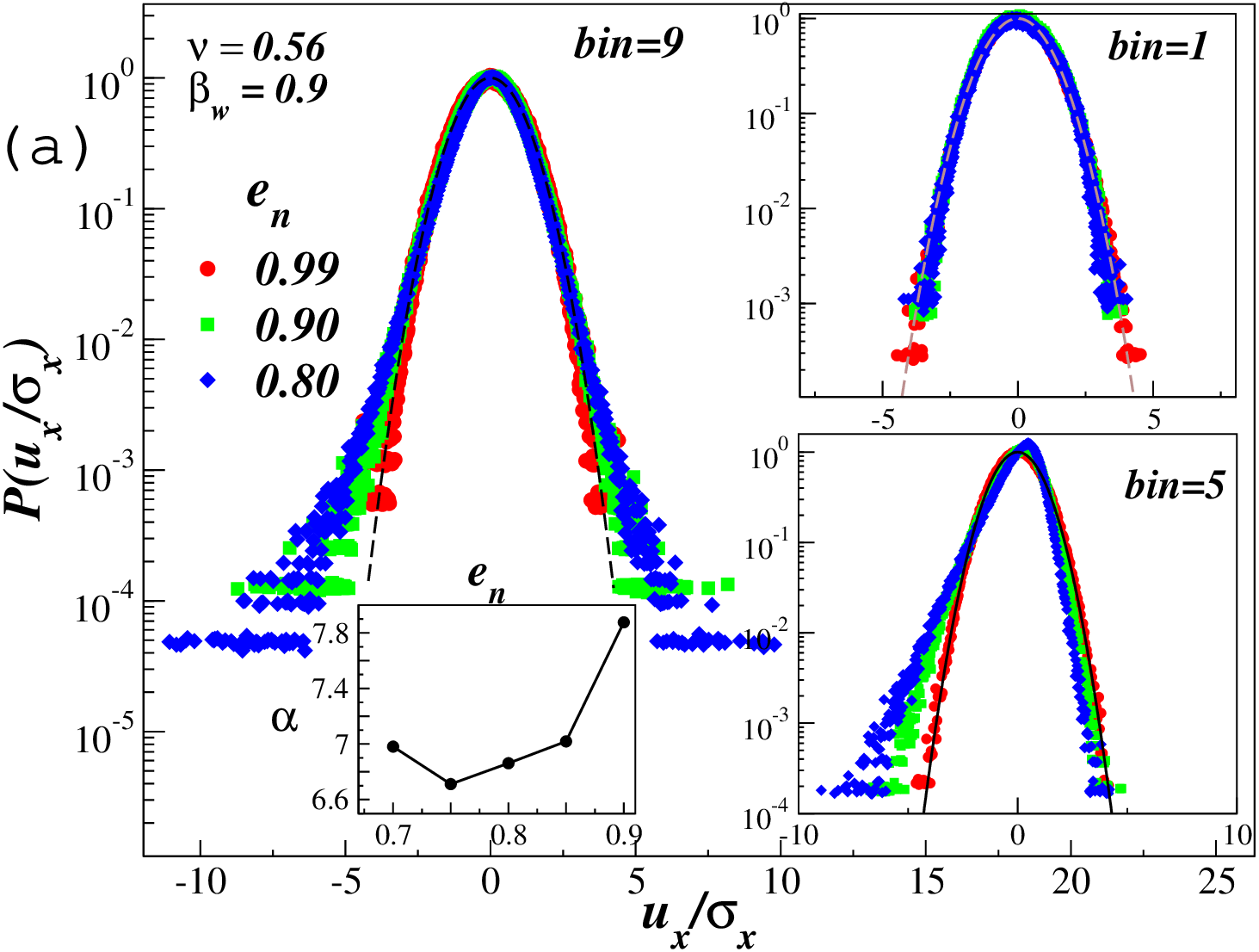}
\end{minipage}
\quad
\begin{minipage}{8.5cm}
\includegraphics[width=0.95\textwidth]{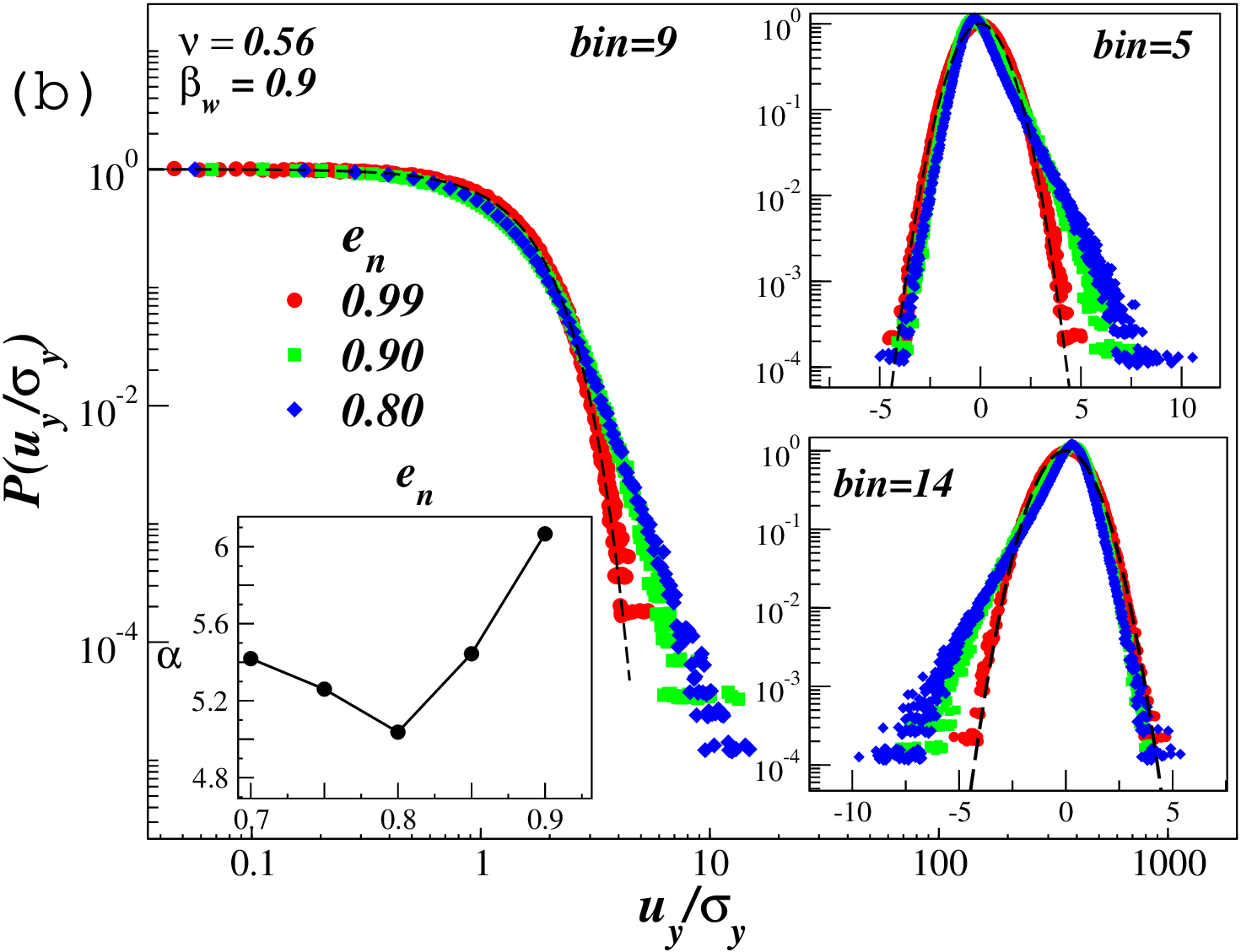}
\end{minipage}
\caption{\small 
(Color online)
(a) Effect of restitution coefficient, $e_n$, on $P(u_x)$ at $\nu=0.56$ for smooth
walls ($\beta_w=0.9$). The velocity distribution in $bin=1$, upper inset,
remains a Gaussian, the distribution in $bin=5$, lower inset, develops
asymmetric tails and the distribution in $bin=9$ makes a transition from 
Gaussian to a power-law tails with increasing dissipation.
Left Inset: Variation of power-law exponent, $\alpha$, with $e_n$.
(b) Same as in panel ($a$) but for $P(u_y)$ in log-log scale
to discern the power-law behviour.
}
\label{fig2}
\end{figure}

For a dense flow, $\nu=0.56$, with  smooth walls $\beta_w=0.9$,
Fig.~\ref{fig2}(a) shows the effect of collisional dissipation, $e_n$, 
on the form of $P(u_x)$ distribution in different bins.
The velocity distribution in $bin=1$ (adjacent to wall), the upper inset in Fig.~\ref{fig2}(a),
remains close to a Gaussian, irrespective of the value of $e_n$.
The $P(u_x)$ distribution in $bin=5$
(between the wall and the channel-centerline), the lower inset in 
Fig.~\ref{fig2}(a), becomes {\it asymmetric} with increasing dissipation,
with an enhanced probability of negative velocities (negative skewness).
At the center of the channel ($bin=9$), the high velocity tails of the distribution function
undergo a transition from a Gaussian to a non-Gaussian (with over-populated tails)
with increasing collisional dissipation, as seen in the main panel of Fig.~2(a).
The analogue of Fig.~2(a) for the fluctuating transverse velocity $u_y$
is shown in Fig.~\ref{fig2}(b).
The transition of $P(u_y)$ in the channel-centerline 
is very similar to that seen in $P(u_x)$.
The distribution in $bin=1$ remains Gaussian and is not shown here,
instead we include the distribution in $bin=14$. 
Note that $P(u_y)$ distribution in $bin=5$ and $bin=14$  are mirror images
since these bins are symmetrically located about the channel-centerline.
The appearance of asymmetric distribution functions in two shear-layers
(with increasing dissipation-level) could be a consequence of 
density-waves~\cite{Gollub01} in narrow shear-layers. 
(Within the plug-region, however, the local distribution functions
are slightly affected by such asymmetries.)
This issue is relegated to a future study.

\begin{figure}[h]
\begin{minipage}[h]{8.5cm}
\includegraphics[width=0.95\textwidth]{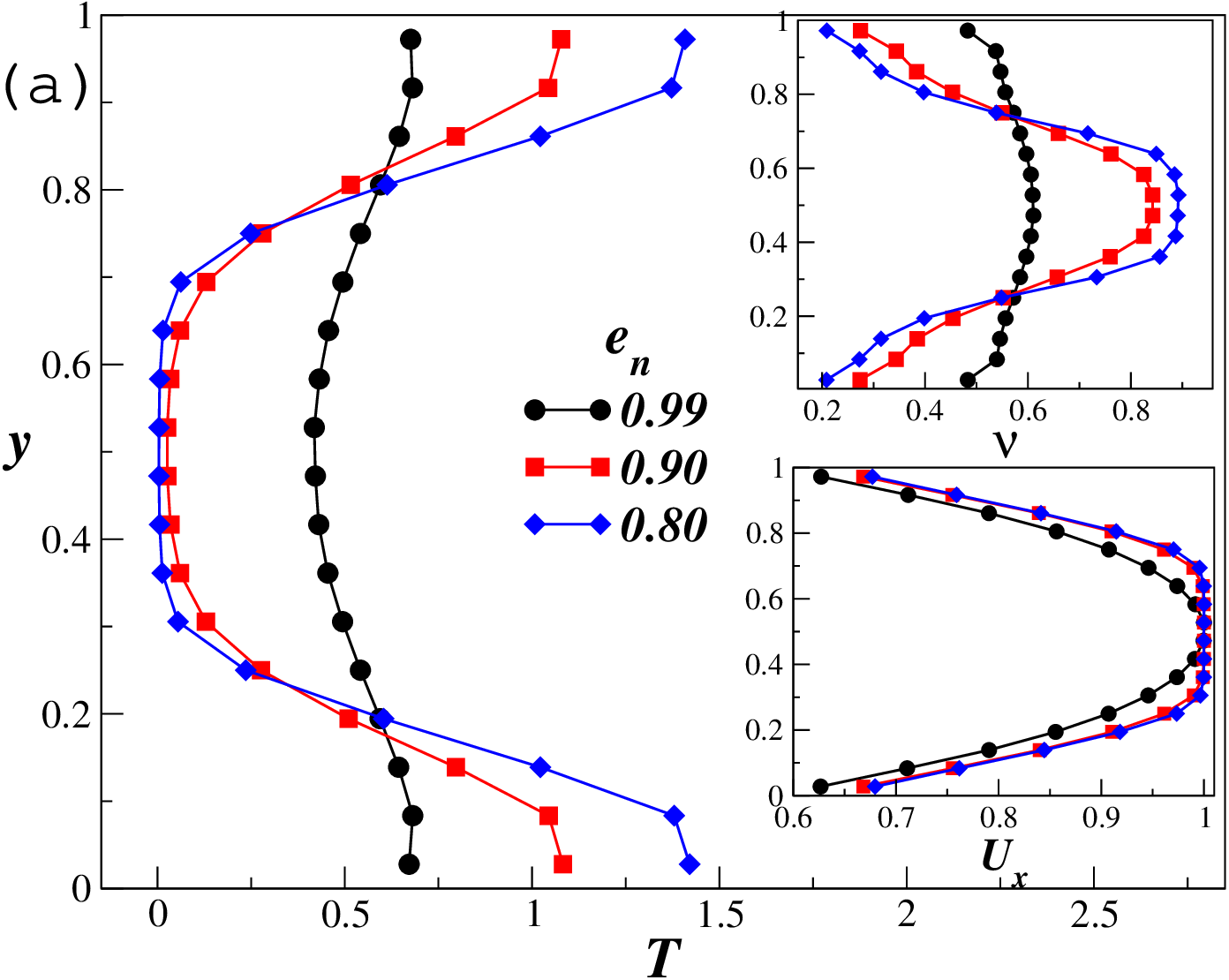}
\end{minipage}
\quad
\begin{minipage}[h]{8.5cm}
\includegraphics[width=0.95\textwidth]{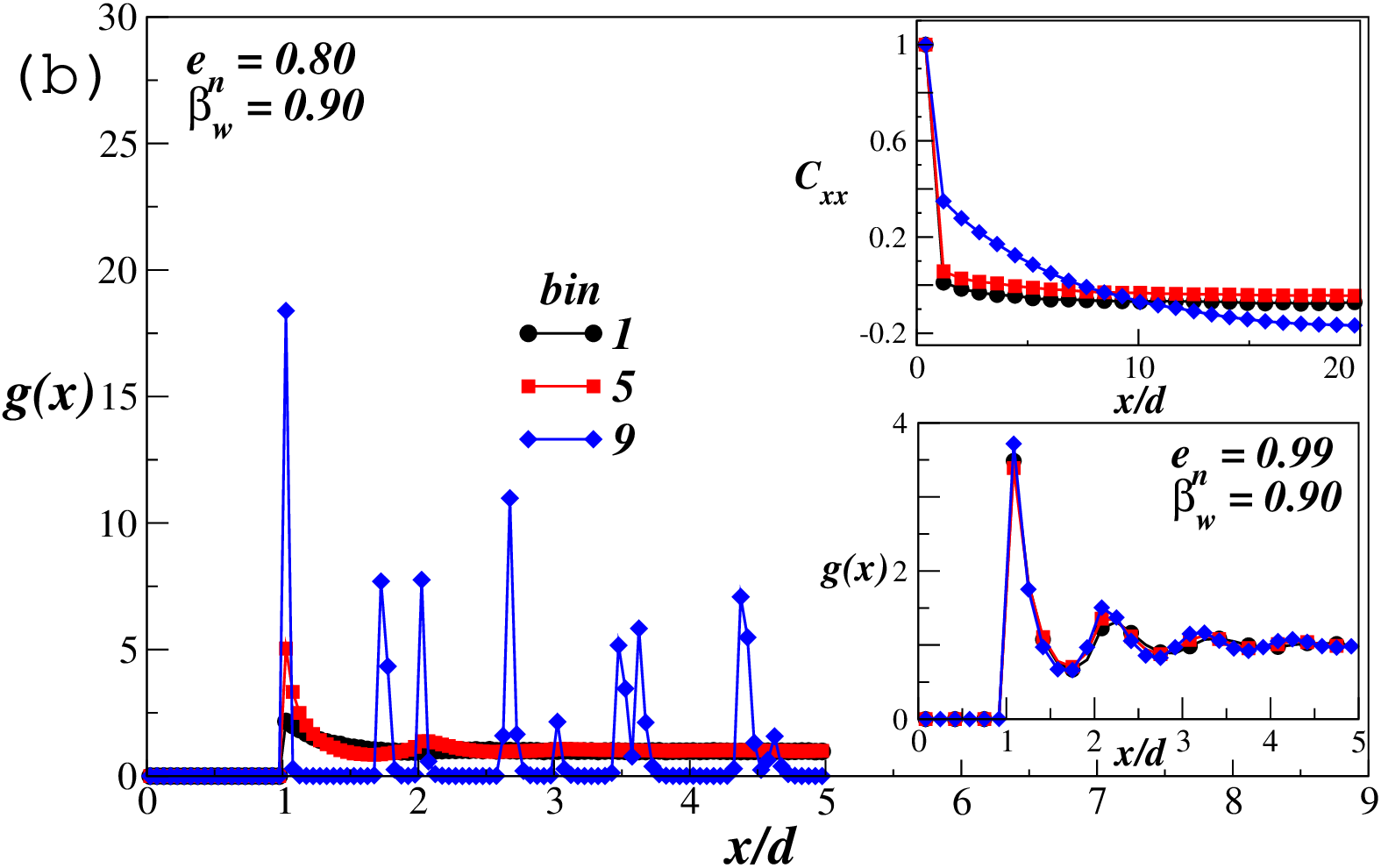}
\end{minipage}
\caption{\small
(Color online)
(a) Variations of $U_x$, $T$ and $\nu$ along $y$, with parameter values as in Fig.~2. 
(b) Main Panel: Pair correlation function, $g(x)$, in different bins
at $\nu=0.56$ for $e_n=0.8$ with smooth walls ($\beta_w=0.9$).
Upper Inset: Streamwise velocity correlation function ($C_{xx}$) with 
parameter values as in main panel.
Lower Inset: $g(x)$ in different bins for $e_n=0.99$ which shows a liquid-like structure.
}
\label{fig3}
\end{figure}

For parameter values as in Fig.~2, 
the  profiles of the mean field quantities are displayed in Fig.~\ref{fig3}(a)
which shows the emergence of a `plug' around the channel-centerline (with negligible
gradients in $U_x$, $T$ and $\nu$) with decreasing $e_n$, 
and two `shear-layers' adjacent to two-walls with steep gradients in $U_x$, $T$ and $\nu$.
To pinpoint the role of $e_n$ on distribution functions, first we study its effect on the 
pair correlation function and the spatial velocity correlations. 
The velocity correlation function is defined as follows:
$$
  C_{ij}=\frac{\left<u_i(x) u_j(x+\delta x)\right>}{\left<u_i(0)u_i(0)\right>},
$$
where the indices $i$,$j$ denote the coordinate directions. 
In the quasielastic limit, the pair correlation function, 
the lower inset in Fig.~\ref{fig3}(b), shows a liquid-like structure in all bins 
and the velocity correlation is close to zero (not shown). 
The signatures of plug-formation with increasing dissipation
show up in the pair correlation function (the main panel of Fig.~\ref{fig3}b) which indicates 
a transition from a liquid to a crystal-like structure in $bin=9$ at $e_n=0.8$.
With increasing density correlation in $bin=9$, the velocity correlation 
$C_{xx}$ also becomes strong as shown in the upper inset of Fig.~\ref{fig3}(b).
(It is interesting to note that the velocity correlation is negative beyond a 
certain correlation length, $x/d\sim  10$, which is an indicator
of circulatory-type motion~\cite{MS98} for the fluctuating velocity field.) 
At $e_n=0.8$, the pair correlation function outside the plug region ($bin=1, 5$)
shows a gaseous structure and the $C_{xx}$ correlation is weak/absent 
(see upper inset in Fig.~\ref{fig3}b). 
Clearly, the enhanced density and velocity correlations 
around the channel-centerline are responsible for the
emergence of non-Gaussian tails with increasing dissipation (Fig.~2).

With reference to dense flows in Fig.~2,
the tails of $P(u_i)$ in the plug-region 
can be fitted via a power-law  of the form $P(u_i)\sim u_i^{-\alpha_i}$,
with a single exponent, $\alpha_x \sim 7$
and $\alpha_y \sim 5.5$, for a range of restitution coefficients,
see the left insets in Figs.~2(a) and 2(b).
The near-constancy of $\alpha_i$ for $e_n<0.85$ is  due to the fact that
the density within the plug saturates to the close-packing limit ($\nu_c\sim 0.9$)
for $e_n<0.85$ and consequently the other hydrodynamic fields also remain
invariant there with a further decrease in $e_n$.
This weak-variation of $\alpha_i$ on $e_n$ is also implicated
in its variation with average density.
A similar power-law behavior (with $\alpha\sim 2.9-7.4$) 
has recently been observed in experiments of gravity-driven channel flow~\cite{MN05};
however, it is difficult to make a direct comparison 
since the experimental geometry is different (with a sieve at the bottom)
and the flow corresponds to the dense `quasi-static' regime.
This issue can be resolved by probing very dense systems which
is beyond the scope of the present paper.

\begin{figure}
\begin{minipage}{9.0cm}
\includegraphics[width=0.95\textwidth]{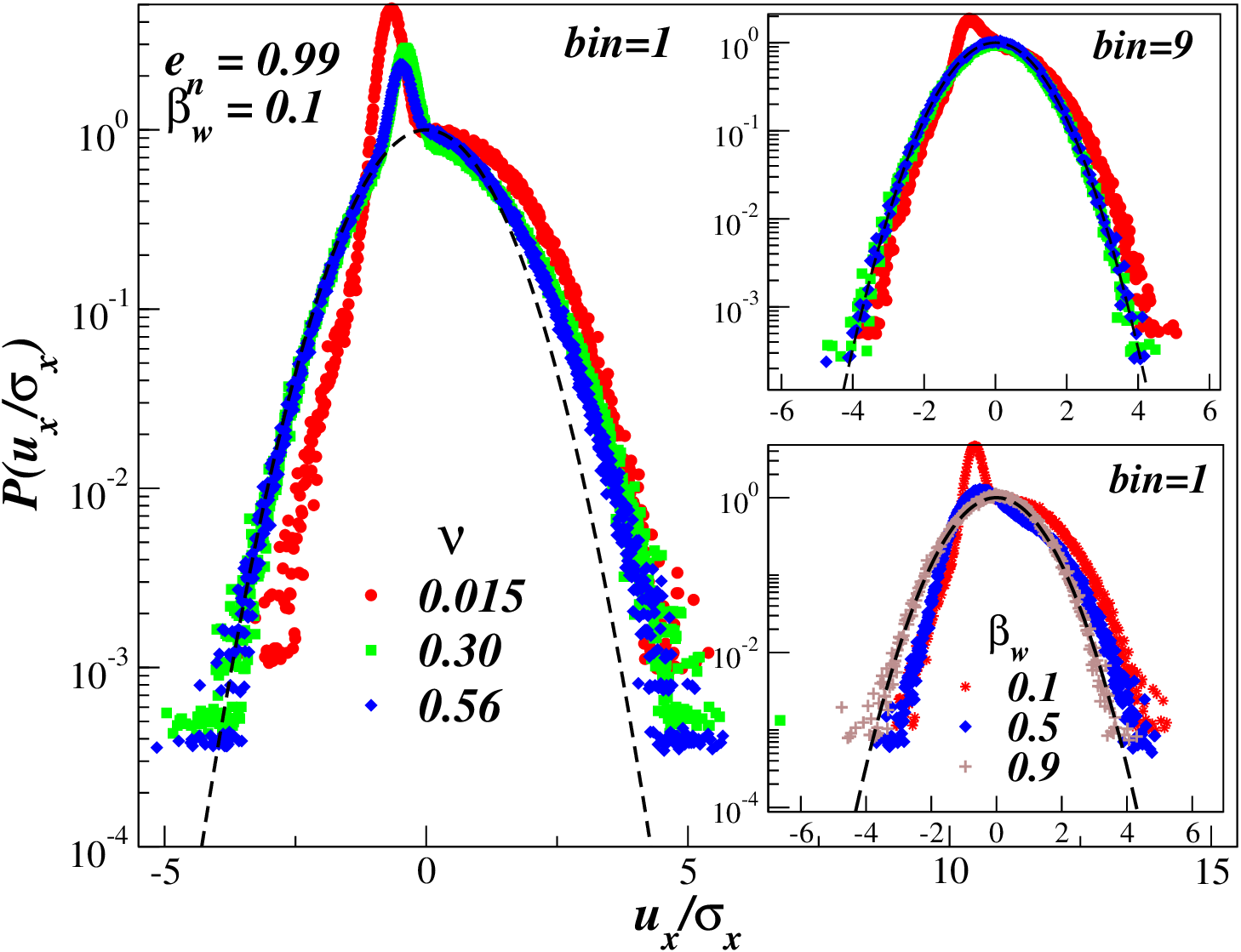}
\end{minipage}
\caption{\small
(Color online)
Effect of mean volume fraction on  $P(u_x)$ 
for a rough-wall ($\beta_w=0.1$) with $e_n=0.99$ in $bin=1$. 
Upper Inset: $bin=9$.
Lower Inset: Effect of $\beta_w$ on $P(u_x)$  at $\nu=0.015$ in $bin=1$.
}
\label{fig4}
\end{figure}

The effect of wall-roughness on $P(u_x)$  is shown in Fig.~\ref{fig4},
for a rough wall $\beta_w=0.1$ with quasielastic collisions $e_n=0.99$. 
The $P(u_x)$ distribution near the wall ($bin=1$, the main panel)
develops a single-peak asymmetric structure at all densities, with its 
peak being positioned at some negative velocity. 
Near the centerline ($bin=9$, upper inset), however,
$P(u_x)$ remains asymmetric only at low densities and becomes a Gaussian at larger densities. 
The corresponding $P(u_y)$ distribution (not shown) remains a Gaussian at all densities
in the quasielastic limit.
For dense flows with {\it rough} walls, both $P(u_x)$ and $P(u_y)$ develop power-law tails
with decreasing $e_n$ around the channel-centerline as in Fig.~2 for smooth walls;
the wall-roughness did not influence the associated power-law exponents.

Focusing on the dilute regime ($\nu=0.015$), the effect of $\beta_w$ on
$P(u_x)$ in $bin=1$ is shown in the lower inset of Figs.~\ref{fig4}.
It is clear that the asymmetry in  $P(u_x)$ diminishes with 
increasing wall smoothness ($\beta_w$) and 
becomes a Gaussian when the walls are smooth ($\beta_w=0.9$).
This wall-roughness-induced asymmetry in $P(u_x)$ is also reflected in the probability 
distribution of the instantaneous particle streamwise velocity $c_x$ (not shown), 
with a single peak in the low velocity region for $bin=1$--
also, $P(c_x)$ for $\nu=0.015$ 
approaches a Gaussian with increased wall-smoothness.
Since the particles loose 
significant amount of tangential velocity while colliding with a {\it rough} 
wall in comparison with their collisions with a {\it smooth} wall,  a peak 
near the low velocity region is expected for rough walls. The greater 
the loss of tangential velocity at walls, the more is the deviation from a Gaussian, 
and the related asymmetry in Fig.~\ref{fig4} is, therefore, tied to wall-roughness.
On the whole, in dilute flows the effect 
of wall-roughness is felt on the local distribution functions throughout the channel,
whereas for dense flows only the near-wall region is affected.

In conclusion, the local velocity distribution functions in a granular Poiseuille 
flow (GPF) with {\it smooth} walls remains Gaussian for a wide range of densities 
for nearly elastic collisions ($e_n\to 1$) which,
in turn, suggests that the GPF (with smooth walls) could serve as a prototype 
`non-equilibrium steady state' 
to derive constitutive models starting from the Boltzmann-Engkog equation.
For dense flows, enhanced density correlations and the related
spatial velocity correlations are responsible
for the emergence of power-law tails with increasing collisional dissipation
around the channel-centerline (irrespective of wall roughness) where the flow undergoes
a transition from a liquid-like to a crystal-like (`plug') structure in the same limit. 
For a rough wall, the near-wall distribution functions
are significantly different from those in the bulk at all densities
which calls for a relook at the 
derivation of granular boundary conditions~\cite{Richman88}.

\end{document}